\documentclass[aps,prd,amsmath,amssymb,preprintnumbers,nofootinbib,a4paper,preprint]{revtex4-1}
\pdfoutput=1
\usepackage{amsthm}
\usepackage{graphicx}
\usepackage{color}
\newcommand{\beq}{\begin{eqnarray}}
\newcommand{\eeq}{\end{eqnarray}}

\newcommand{\bmp}{\noindent\begin{minipage}{16cm}}
\newcommand{\emp}{\end{minipage}\vskip 7mm} 

\usepackage{dcolumn}
\usepackage{bm}
\usepackage{bbm}
\usepackage{subfigure}
\usepackage{pxfonts}

\theoremstyle{definition}

\theoremstyle{plain}

\usepackage{epsfig}

\usepackage[ margin=5pt, font=normalsize,labelfont=bf,justification=raggedright]{caption}
\usepackage{youngtab}
\usepackage{slashed}

\definecolor{rossoCP3}{cmyk}{0,.88,.77,.40}

\def\lsim{\mathrel{\rlap{\lower4pt\hbox{\hskip1pt$\sim$}}
    \raise1pt\hbox{$<$}}}                
\def\gsim{\mathrel{\rlap{\lower4pt\hbox{\hskip1pt$\sim$}}
    \raise1pt\hbox{$>$}}}                

\begin{document}
\title{\Large  \color{rossoCP3}   
Light Asymmetric Dark Matter on the Lattice:\\ SU(2) Technicolor with Two Fundamental Flavors  \\ ~ }
\author{Randy Lewis$^{\color{rossoCP3}{\spadesuit}}$}\email{randy.lewis@yorku.ca} 
\author{Claudio Pica$^{\color{rossoCP3}{\varheartsuit}}$}\email{pica@cp3-origins.net} 
\author{Francesco Sannino$^{\color{rossoCP3}{\varheartsuit}}$}\email{sannino@cp3-origins.net} 
 \affiliation{
$^{\color{rossoCP3}{\spadesuit}}$Department of Physics and Astronomy,  {\color{rossoCP3} York University}, Toronto, M3J 1P3, Canada
}
 \affiliation{
$^{\color{rossoCP3}{\varheartsuit}}${ \color{rossoCP3}  \rm CP}$^{\color{rossoCP3} \bf 3}${\color{rossoCP3}\rm-Origins} \& the Danish Institute for Advanced Study {\color{rossoCP3} \rm DIAS},\\ 
University of Southern Denmark, Campusvej 55, DK-5230 Odense M, Denmark
\vspace{1cm}
}
 
\begin{abstract}
The SU(2) gauge theory with two massless Dirac flavors constitutes the building block of several models of Technicolor. Furthermore it has also been used as a template for the construction of a natural light asymmetric, or mixed type, dark matter candidate. We use explicit lattice simulations to confirm the pattern of chiral symmetry breaking by determining the Goldstone spectrum and therefore show that the dark matter candidate can, de facto, be constituted by a complex Goldstone boson.  We also determine the phenomenologically relevant spin-one and spin-zero isovector spectrum and demonstrate that it is well separated from the Goldstone spectrum. 
 \\ ~ \\
[.1cm]
{\footnotesize  \it Preprint: CP$^3$-Origins-2011-28 \& DIAS-2011-15}
 \end{abstract}

\maketitle

\section{Introduction}
The nature of dark matter (DM) is an important problem in modern physics.
DM plays a key role in the formation of large structures and the evolution of
the Universe. It is also widely expected to provide a link to physics beyond
the Standard Model (SM).  For these reasons much experimental, observational,
and theoretical effort has been devoted to shedding light on DM.
It is
popular to identify DM with Weakly Interacting Massive Particles (WIMPs).

Many properties of a WIMP are not constrained by our current knowledge of DM,
for example the WIMP may or may not be a stable particle~\cite{Nardi:2008ix} and
it may or may not be identified with its antiparticle~\cite{Nussinov:1985xr}.
Asymmetric DM refers to a scenario where the WIMP's antiparticle 
has been annihilated away, leaving the WIMP itself as the observed DM.
Asymmetric DM candidates were put forward in \cite{Nussinov:1985xr} as
technibaryons, in \cite{Gudnason:2006ug} as Goldstone bosons, and subsequently
in many diverse forms \cite{Foadi:2008qv,Khlopov:2008ty,Dietrich:2006cm,Sannino:2009za,Ryttov:2008xe,Kaplan:2009ag,Frandsen:2009mi}.
There is also the possibility of mixed DM~\cite{Belyaev:2010kp}, i.e.\ having
both a thermally-produced symmetric component and an asymmetric one.

Null results from several experiments, such as CDMS~\cite{Ahmed:2010wy} and
Xenon10/100~\cite{Angle:2011th,Aprile:2011hi},
have placed stringent constraints on WIMP-nucleon cross sections.
Interestingly DAMA~\cite{Bernabei:2008yi} and CoGeNT~\cite{Aalseth:2010vx} have
both produced evidence for an annual modulation signature for DM, as
expected due to the relative motion of the Earth with respect to the DM halo.
These results support a light WIMP with mass of order a few GeV, which offers
the attractive possibility of a common mechanism for baryogenesis
and DM production. At first glance it seems that the WIMP-nucleon cross
sections required by DAMA and CoGeNT have been excluded by CDMS and Xenon upon
assuming spin-independent interactions between WIMPs and nuclei (with protons
and neutrons coupling  similarly to WIMPs), but a number of resolutions for
this puzzle have been proposed in the literature~\cite{Khlopov:2010pq,TuckerSmith:2001hy,Chang:2010yk,Feng:2011vu,Frandsen:2011ts,DelNobile:2011je}. Interestingly, new results from  the CRESST-II experiment report signals of light DM \cite{Angloher:2011uu}. 

A composite origin of DM is an intriguing possibility given that the
bright side of the universe, constituted mostly by nucleons, is also composite.
Thus a new strongly-coupled theory could be at the heart of DM.
In this work we investigate, for the first time on the lattice, a
technicolor-type extension of the SM expected to naturally yield a light DM candidate, as introduced in \cite{Ryttov:2008xe} and used in
\cite{DelNobile:2011je} to reconcile the experimental observations.

\section{A Light Dark Matter Template: $SU(2)$ Technicolor with two Dirac Flavors}
Asymmetric DM requires at least one complex field with a (nearly)
conserved Abelian quantum number. It would be exciting if the theory predicting
the existence of this state is also directly involved with the breaking of
electroweak symmetry. Therefore a natural candidate is a technicolor model.
Most of the states of the theory are much heavier than a few GeV, but if there
are Goldstone bosons not absorbed by the longitudinal degrees of freedom of the
SM massive gauge bosons, they become primary candidates for producing a
natural hierarchy between the GeV and the TeV scale. 

The minimal technicolor theory that breaks the electroweak symmetry and
features a light DM state was constructed first in
\cite{Appelquist:1999dq,Duan:2000dy}.
This theory is $SU(2)$ technicolor with two Dirac flavors,
which has global symmetry $SU(4)$ expected to break to $Sp(4)$.
Five Goldstone bosons are generated of which three are eaten by the SM gauge
bosons and a complex one (a techni-diquark which is essentially a technibaryon
of this $SU(2)$ theory) remains in the spectrum.
This is our candidate for the light asymmetric DM particle.

The walking version of this model,
known as Ultra Minimal Walking Technicolor (UMT),
has been constructed in \cite{Ryttov:2008xe} and it contains,
besides the fermions in the fundamental representation, also a Dirac fermion in
the adjoint.  For the present exploratory work, we begin with Section III of
\cite{Ryttov:2008xe} but without the adjoint fermion.
Moreover, when implementing our lattice simulations we make two additional
modifications: all
electroweak interactions are omitted from the lattice simulations because that
physics is well-understood by perturbative methods, and explicit technifermion
masses are added because
lattice simulations with exactly massless fermions are not practical.
The resulting Lagrangian is very simple
\begin{equation}\label{Lagrangian}
\mathcal{L} =  -\frac{1}{4}{F}_{\mu\nu}^a F^{a\mu\nu}
               + \overline{U}(i\gamma^{\mu}D_{\mu}-m)U
               + \overline{D}(i\gamma^{\mu}D_{\mu}-m)D \ ,
\end{equation}
where $U$ and $D$ are the two techniquark fields having a common bare mass
$m$, $F_{\mu\nu}^a$ is the field strength, and $D_\mu$ is the covariant
derivative.  The Dirac and technicolor indices of $U$ and $D$ are not
shown explicitly.

Lattice simulations will be used to extrapolate to $m=0$, and
in that limit the Lagrangian has a global $SU(4)$ symmetry corresponding to
the four chiral fermion fields
\begin{equation}
U_L=\frac{1}{2}(1-\gamma^5)U \ ,
~~~~~
U_R=\frac{1}{2}(1+\gamma^5)U \ ,
~~~~~
D_L=\frac{1}{2}(1-\gamma^5)D \ ,
~~~~~
D_R=\frac{1}{2}(1+\gamma^5)D \ .
\end{equation}
For $m\neq0$, the $SU(4)$ symmetry is explicitly broken to a remaining $Sp(4)$
subgroup as follows.  The Lagrangian from (\ref{Lagrangian}) can be rewritten as
\begin{equation}\label{LagrangianQ}
\mathcal{L} =  -\frac{1}{4}{F}_{\mu\nu}^a F^{a\mu\nu}
               + i\overline{U}\gamma^{\mu}D_{\mu}U
               + i\overline{D}\gamma^{\mu}D_{\mu}D
               + \frac{m}{2}Q^T (-i\sigma^2) C \,EQ + { \frac{m}{2}\left(Q^T(-i\sigma^2)C\,EQ \right)^{\dagger} } \ ,
\end{equation}
where
\begin{equation}
Q = \left( \begin{array}{c} U_L \\ D_L \\ -i\sigma^2C\overline{U}_R^T \\
    -i\sigma^2C\overline{D}_R^T \end{array} \right) \ ,
~~~~~~~~
E = \left(\begin{array}{cccc} 0 & 0 & 1 & 0 \\
      0 & 0 & 0 & 1 \\ -1 & 0 & 0 & 0 \\ 0 & -1 & 0 & 0 \end{array}\right) \ ,
\end{equation}
$C$ is the charge conjugation operator acting on Dirac indices, and
the Pauli structure $-i\sigma^2$ is the standard antisymmetric tensor acting on
color indices.
Under an infinitesimal $SU(4)$ transformation defined by
\begin{equation}
Q \to \left(1+i\sum_{n=1}^{15}\alpha^nT^n\right)Q \ ,
\end{equation}
the Lagrangian (\ref{LagrangianQ}) becomes
\begin{equation}
\mathcal{L} \to \mathcal{L} + \frac{im}{2}\sum_{n=1}^{15}\alpha^nQ^T (-i\sigma^2) C \,
                              \left(ET^n+T^{nT}E\right)Q + {{\rm h.c.}}\ ,
\end{equation}
where $T^n$ denotes the 15 generators of $SU(4)$ and $\alpha^n$ is a set of
15 constants.
The only generators that leave the Lagrangian invariant are those that obey
\begin{equation}
ET^n + T^{nT}E = 0
\end{equation}
which is precisely the definition of an $Sp(4)$ Lie algebra.  From this, it is
straightforward to derive the ten $Sp(4)$ generators in a specific
representation; see the appendix of \cite{Ryttov:2008xe}.

For $m=0$ the Lagrangian retains the full $SU(4)$ symmetry but,
by analogy with the $SU(3)$ theory of QCD, one might expect dynamical symmetry
breaking associated with the appearance of a nonzero vacuum expectation value,
\begin{eqnarray}
\langle \overline{U}U + \overline{D}D \rangle \neq0 \ .
\end{eqnarray}
Since this vacuum expectation value has the same structure as the terms
containing $m$ in the Lagrangian, the dynamical breaking would also be
$SU(4)\to Sp(4)$.
According to Noether's theorem, the five broken generators would be accompanied
by five Goldstone bosons.

Of course this suggestion of dynamical symmetry breaking must be checked
nonperturbatively using first-principles lattice simulations.
As reported below, we have done so and our lattice simulations provide direct
verification of this dynamical symmetry breaking.

There have been several previous lattice studies of $SU(2)$ gauge theory with
fermions in the fundamental representation~\cite{Hands:1999md}, mainly
motivated by interest at nonzero chemical potential, but all of these studies
relied on the staggered action where the number of fermions is a multiple of
4.  Our minimal technicolor theory requires $N_f=2$ and thus we use the
Wilson action instead of staggered fermions.  When studying chiral symmetry
breaking scenarios with Wilson fermions attention must be paid to the
presence, on a lattice, of the unphysical Aoki phase.  For fixed gauge
coupling, the Aoki phase is entered as the quark mass is reduced.  An
analytic discussion of the Aoki phase symmetries for this theory is provided
in \cite{DelDebbio:2008wb}, and three lattice studies are also available
\cite{Fukushima:2008su,Matsufuru:2009zz,Nagai:2009ip}.  For our present
simulations, we
avoid the Aoki phase and work exclusively in the physical phase.  There have
been very few previous lattice results reported for $SU(2)$ gauge theory with
2 fundamental fermions \cite{Nagai:2009ip,Catterall:2007yx,Hietanen:2008mr}
and in each case the primary focus was on a different action (either $N_f>2$
or adjoint representation fermions).  Our work is the first lattice study
focused on the mass spectrum of the two-color two-flavor theory, which is
the familiar technicolor template.

\section{Lattice Hadron Operators and Effective Field Theory}\label{sec:EFT}

The creation operator for a meson is the Hermitian conjugate of its annihilation
operator, and a set of local annihilation operators for mesons is
\begin{eqnarray}
\mathcal{O}_{\overline{U}D}^{(\Gamma)}
&\equiv& \overline{U}(x)\Gamma D(x) \ , \nonumber \\
\mathcal{O}_{\overline{D}U}^{(\Gamma)}
&\equiv& \overline{D}(x)\Gamma U(x) \ , \nonumber \\
\mathcal{O}_{\overline{U}U\pm\overline{D}D}^{(\Gamma)}
&\equiv& \frac{1}{\sqrt{2}}
\bigg(\overline{U}(x)\Gamma U(x) \pm \overline{D}(x)\Gamma D(x)\bigg) \ ,
\label{GBmesonops}
\end{eqnarray}
where $\Gamma$ is a chosen Dirac structure.  In this work we consider
$\Gamma=1$, $\gamma^5$, $\gamma^\mu$, or $\gamma^\mu\gamma^5$.
In lattice simulations, meson masses are extracted from the time dependence
of correlation functions, for example
\begin{eqnarray}
C_{\overline{U}D}^{(\Gamma)}(t_x-t_y)
&=& \sum_{\vec x}\sum_{\vec y}\mathcal{O}_{\overline{U}D}^{(\Gamma)}(y)
    \left(\mathcal{O}_{\overline{U}D}^{(\Gamma)}(x)\right)^{\dagger} \nonumber \\
&=& \sum_{\vec x}\sum_{\vec y}{\rm Tr}\left[\Gamma D(y)\overline{D}(x)\gamma^0
    \Gamma^\dagger\gamma^0U(x)\overline{U}(y)\right] \, ,
\end{eqnarray}
where Tr$[\cdots]$ denotes a trace over Dirac and color indices and we have dropped the vacuum expectation values for the propagators as explained in Appendix \ref{Appendix1}.

Perhaps surprisingly, local diquark (i.e.\ baryon) correlation functions
provide no new data in this theory.  To understand why, notice that the
available local diquark operators are
\begin{eqnarray}
\mathcal{O}_{UD}^{(\Gamma)}
&\equiv& U^T(x)(-i\sigma^2)C\Gamma D(x) \ , \nonumber \\
\mathcal{O}_{DU}^{(\Gamma)}
&\equiv& D^T(x)(-i\sigma^2)C\Gamma U(x) \ .
\label{GBdiquarkops}
\end{eqnarray}
(Some operators containing $U^T\cdots U$ or $D^T\cdots D$ are identically zero.)
The diquark correlation function is therefore
\begin{eqnarray}
C_{UD}^{(\Gamma)}(t_x-t_y)
&=& \sum_{\vec x}\sum_{\vec y}\mathcal{O}_{UD}^{(\Gamma)}(y)
    \left(\mathcal{O}_{UD}^{(\Gamma)}(x)\right)^{\dagger} \nonumber \\
&=& \sum_{\vec x}\sum_{\vec y}{\rm Tr}\left[\Gamma D(y)\overline{D}(x)\gamma^0
    \Gamma^\dagger C^\dagger(-i\sigma^2)^\dagger\gamma^{0T}\overline{U}^T(x)
    U^T(y)(-i\sigma^2)C\right]
\end{eqnarray}
and this can be rewritten by using two properties of the charge conjugation
operator: one for a Dirac matrix,
\begin{equation}
\gamma^{\mu T} = -C\gamma^\mu C^\dagger \ ,
\end{equation}
and the other for the Wilson fermion matrix,
\begin{equation}
C^{-1}(-i\sigma^2)^{-1}\bigg(U(y)\overline{U}(x)\bigg)^TC(-i\sigma^2)
= U(x)\overline{U}(y) \ ,
\label{14}
\end{equation}
to arrive at
\begin{eqnarray}
C_{UD}^{(\Gamma)}(t_x-t_y)
&=& \sum_{\vec x}\sum_{\vec y}{\rm Tr}\left[\Gamma D(y)\overline{D}(x)\gamma^0
    \Gamma^\dagger\gamma^0U(x)\overline{U}(y)\right] \nonumber \\
&=& C_{\overline{U}D}^{(\Gamma)}(t_x-t_y)
\end{eqnarray}
for any choice of $\Gamma$.
This conclusion means that a lattice simulation will find
numerically-identical correlation functions, and thus identical masses, for
the meson and diquark.  We have verified this explicitly in the lattice
simulations described below.

Notice that the degenerate pairs have equal
angular momentum but opposite parities, for example
\begin{eqnarray}
J\left(\mathcal{O}_{UD}^{(\Gamma)}\right)
&=& J\left(\mathcal{O}_{\overline{U}D}^{(\Gamma)}\right) \ , \\
P\left(\mathcal{O}_{UD}^{(\Gamma)}\right)
&=& -P\left(\mathcal{O}_{\overline{U}D}^{(\Gamma)}\right) \ .
\end{eqnarray}
This relationship for $J^P$ has also been mentioned, for example,
in \cite{Hands:2007uc}.
We must conclude that the three {\em pseudoscalar} meson
Goldstone bosons are accompanied by two {\em scalar} diquark Goldstone bosons.
This is in contrast to the identification made in \cite{Ryttov:2008xe}, where
it was assumed that all five Goldstones would be pseudoscalars.  Despite this,
the effective Lagrangian in
\cite{Ryttov:2008xe} remains unaltered because all scalar and pseudoscalar
particles were retained in that work with the correct assignment with respect to the broken and unbroken generators of the chiral symmetry group\footnote{In practice the only change is the redefinition of $\widetilde{\Pi}_{UD}$ and $\widetilde{\Pi}_{\overline{UD}}$  at the underlying operator level with $\Pi_{UD}$ and  ${\Pi}_{\overline{UD}}$ in  \cite{Ryttov:2008xe}. }. 

Now that the five Goldstones have been identified, we can define an effective
theory wherein those five are the only fields.  Indeed, our lattice simulations
(discussed below) indicate that the non-Goldstone scalar/pseudoscalar
hadrons are even heavier than the vector mesons, making it quite natural to
integrate all non-Goldstones out of the effective theory.
The annihilation operators are conveniently collected into a compact form,
\begin{equation}
\delta{\mathcal L} =
\sum_n\bigg(Q^T(-i\sigma^2)C\Gamma T^nQ\bigg)\Phi^{(\Gamma)n} \, ,
\end{equation}
where $\Phi^{(\Gamma)n}$ is a generic name for the $n$'th particle with
Dirac structure $\Gamma$.
This $\delta{\mathcal L}$ represents the effective Lagrangian couplings for
external fields $\Phi^{(\Gamma)n}$.
For the special case of the five Goldstone bosons, we sum $n$ only over the
five broken generators and we must choose $\Gamma=\gamma^5$.  The result is
\begin{equation}
\delta{\mathcal L}_{\cal G} = Q^T(-i\sigma^2)C\gamma^5{\cal G}Q
\end{equation}
where the matrix ${\cal G}$ contains the five Goldstone bosons (i.e.
the pseudoscalar mesons $\Pi^+$, $\Pi^-$, $\Pi^0$ and the scalar
diquarks $\Pi_{UD}$, $\Pi_{\overline{U}\overline{D}}$):
\begin{equation}
{\cal G} = \frac{i}{2}\left(\begin{array}{cccc}
 0 & \sqrt{2}\Pi_{UD} & \Pi^0 & \sqrt{2}\Pi^+ \\
 -\sqrt{2}\Pi_{UD} & 0 & \sqrt{2}\Pi^- & -\Pi^0 \\
 -\Pi^0 & -\sqrt{2}\Pi^- & 0 & -\sqrt{2}\Pi_{\overline{U}\overline{D}} \\
 -\sqrt{2}\Pi^+ & \Pi^0 & \sqrt{2}\Pi_{\overline{U}\overline{D}} & 0
\end{array}\right) \ .
\end{equation}
The matrix ${\cal G}$ is closed under $Sp(4)$ transformations: the Goldstone bosons
form a five-dimensional representation of $Sp(4)$.
The matrix ${\cal G}$ is also identical to $M_4$ as defined by
equation (22) of \cite{Ryttov:2008xe} after the non-Goldstone fields are removed
from $M_4$.

To summarize, an effective field theory for the five Goldstone bosons
is obtained by using the matrix ${\cal G}$ in place of $M_4$ in
\cite{Ryttov:2008xe}.  The three pseudoscalar Goldstones ($\Pi^\pm$, $\Pi^0$)
are responsible for electroweak symmetry breaking, and the two scalar
Goldstones
($\Pi_{UD}$, $\Pi_{\overline{U}\overline{D}}$)
are the DM candidate and its antiparticle.

\section{Methods for Numerical Simulations}

\begin{figure}
\includegraphics[width=17cm,trim=0 50 0 0,clip=true]{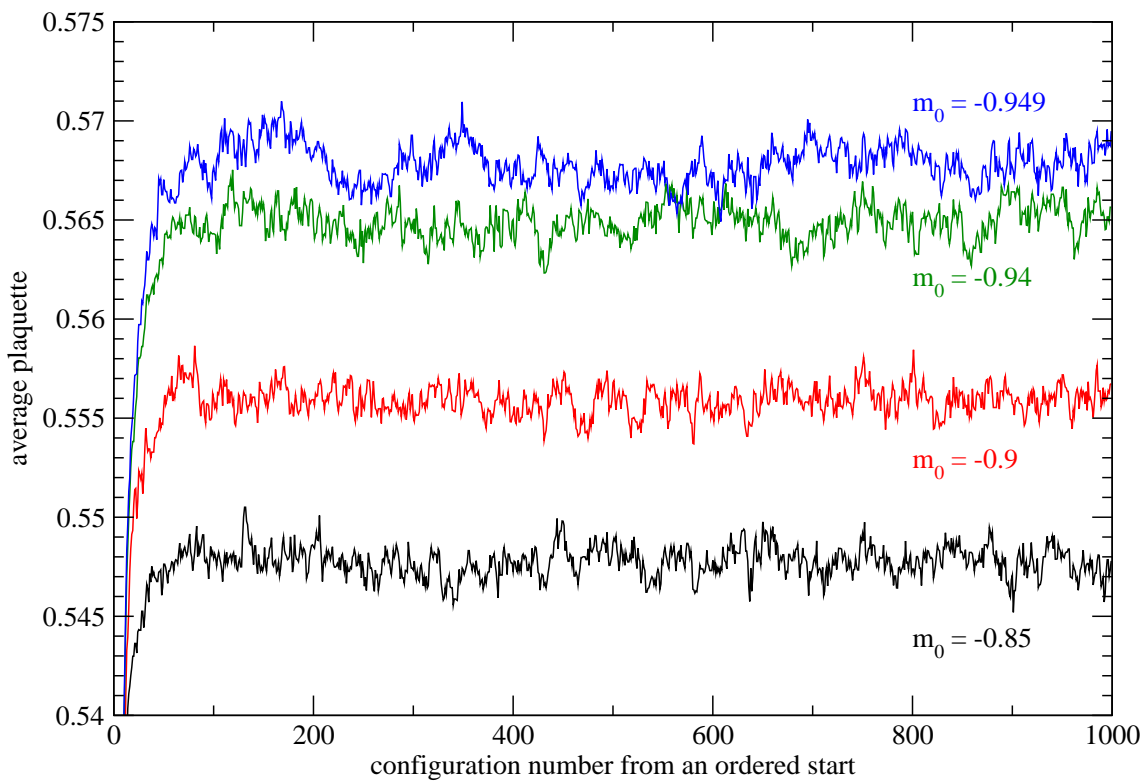}
\caption{Evolution of the average plaquette for selected bare masses at
         $\beta=2.0$.  Our ensemble of 35 configurations is comprised of those
         numbered 320, 340, 360, \ldots 1000.}\label{fig:plaq2p0}
\end{figure}

\begin{figure}
\includegraphics[width=17cm,trim=0 50 0 0,clip=true]{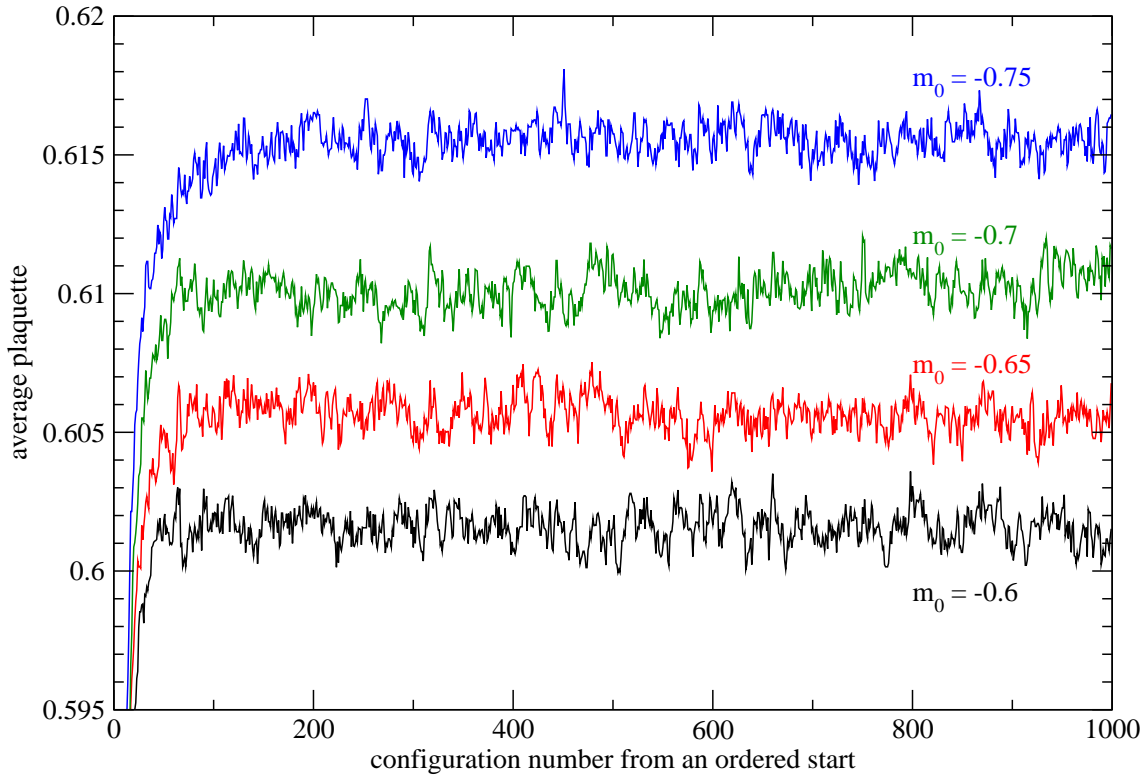}
\caption{Evolution of the average plaquette for selected bare masses at
         $\beta=2.2$.  Our ensemble of 35 configurations is comprised of those
         numbered 320, 340, 360, \ldots 1000.}\label{fig:plaq2p2}
\end{figure}

The standard Wilson action,
\begin{eqnarray}
S_W &=& \frac{\beta}{2}\sum_{x,\mu,\nu}\left(1-\frac{1}{2}{\rm ReTr}
        U_\mu(x)U_\nu(x+\hat\mu)U_\mu^\dagger(x+\hat\nu)U_\nu^\dagger(x)
        \right)
        + (4+m_0)\sum_x\bar\psi(x)\psi(x) \nonumber \\
    & & - \frac{1}{2}\sum_{x,\mu}\bigg(
        \bar\psi(x)(1-\gamma_\mu)U_\mu(x)\psi(x+\hat\mu)
       +\bar\psi(x+\hat\mu)(1+\gamma_\mu)U_\mu^\dagger(x)\psi(x)\bigg) \, ,
       \label{wilsonm}
\end{eqnarray}
is used for this study of $SU(2)$ gauge theory with
two mass-degenerate fermions in the fundamental representation.  Configurations
were generated using the HMC algorithm as implemented in the HiRep
code~\cite{DelDebbio:2008zf}.
A total of 12 ensembles were created, corresponding to six different bare quark
masses $m_0$ for each of two different gauge couplings $\beta$; see
Table \ref{tab:betam0} for details.
We note that \cite{Nagai:2009ip} also contains some simulations of this theory
at $\beta=2.0$ and our findings for the pseudoscalar meson mass and PCAC
quark mass are consistent with that paper.
For other parameter choices, see \cite{Catterall:2007yx} and \cite{Hietanen:2008mr}.
\begin{table}
\caption{Numerical values of the gauge coupling and the bare quark mass used to
generate the 12 ensembles of this project.}
\begin{tabular}{ll}
\hline
$\beta$ & $m_0$ \\
\hline
2.0~~~~~ & -0.85, -0.90, -0.94, -0.945, -0.947, -0.949 \\
2.2 & -0.60, -0.65, -0.68, -0.70, -0.72, -0.75 \\
\hline
\end{tabular}
\label{tab:betam0}
\end{table}
In the present study,
all lattices are $L^3\times T=16^3\times32$ with periodic boundary conditions
in each direction.  Every ensemble contains 35 configurations
separated by 20 unused configurations after an initial thermalization of 320
configurations.
Figures \ref{fig:plaq2p0} and \ref{fig:plaq2p2} display the fluctuations in the
average plaquette within selected Markov chains.

Quark propagators are created from wall sources built of random $U(1)$ phases
at every lattice site on one chosen time step, $t_i$.
Inversions are performed with the standard BiCGstab algorithm.
To reduce statistical fluctuations, correlation functions are averaged over all
lattice times $0\leq t_i\leq31$.
Correlation functions then depend on the time $t_f$ corresponding to the
annihilation operator.  This annihilation operator is local and summed over all
spatial sites to produce a zero-momentum hadron.

Our multi-state fits,
\begin{equation}\label{cosh}
C(t) = \sum_{j=1}^na_j\cosh\left(m_j\left(t-\frac{T}{2}\right)\right) \, ,
\end{equation}
use every nonzero time separation $t=t_f-t_i>0$, which avoids the subjectivity
of choosing a fitting window.
Pseudoscalar, vector and temporal-axial ($A_4$) correlators are fitted to three
states, $n=3$ in (\ref{cosh}); scalar and spatial-axial ($A_1,A_2,A_3$)
correlators use $n=2$.
Statistical uncertainties are produced from each 35-configuration ensemble by
creating 150 bootstrap ensembles (having 35 configurations each).

Isovector hadrons are sufficient for most topics addressed in this work, but
isoscalars are used to study a specific issue.
Isoscalar operators require the computation of single-site propagators,
i.e.\ quark propagators that begin
and end on a single lattice site.  In practice, to obtain a signal requires an
average over many sites, so isoscalar studies are expensive.  Improved methods
have recently been developed \cite{Foley:2005ac,Morningstar:2011ka},
but the needs of our present project are fulfilled
by a simpler tactic.  We choose our ``most physical'' ensemble
($\beta=2.2$ and $m_0=-0.75$) and calculate 2048 single-site propagators, that
is 64 single-site propagators spread uniformly across each lattice time step.

\begin{figure}
\includegraphics[width=17cm,trim=0 50 0 0,clip=true]{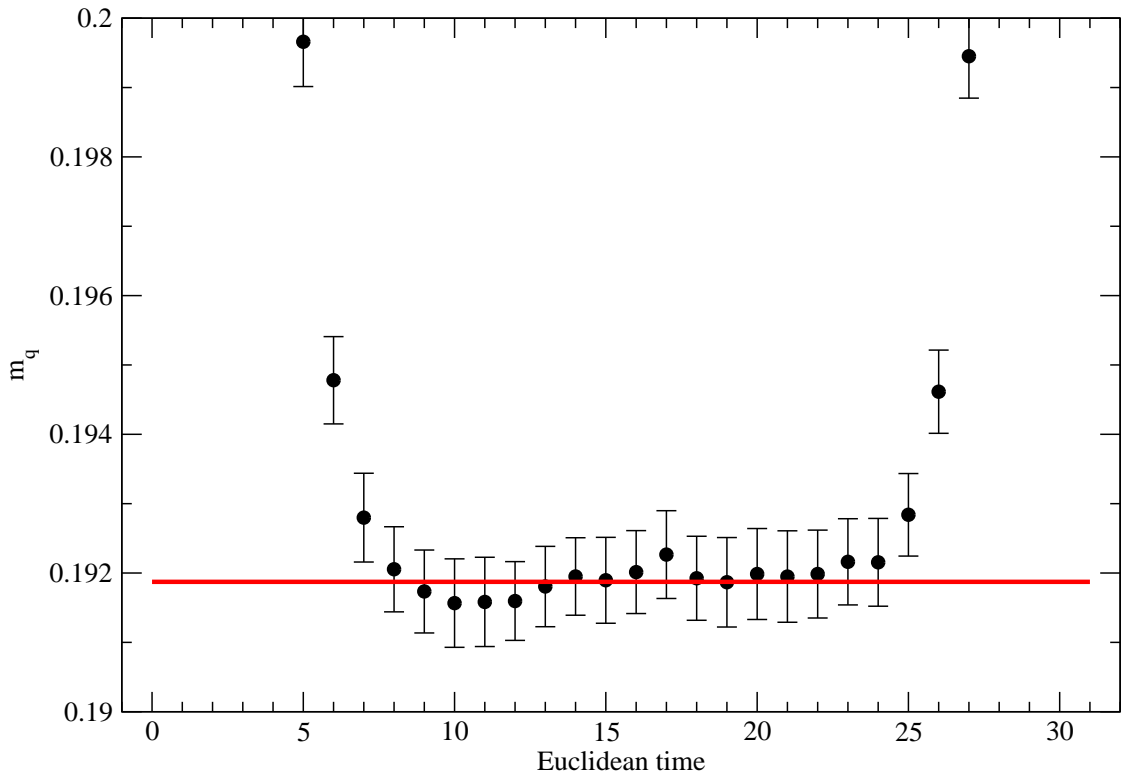}
\caption{The $t$ dependence from the right-hand side of
         (\protect\ref{mqformula}) for the case of $\beta=2.2$ and $m_0=-0.75$.
         The horizontal line is a fit to Euclidean times between 10 and 22
         inclusive.}\label{fig:pcacfit}
\end{figure}

Two additional quantities that can be derived from two-point correlation
functions are also valuable for the present work.  One is the PCAC quark mass
defined by
\begin{equation}\label{mqformula}
m_q = \lim_{t\to\infty}\left(\frac{\langle A_4(t+1)P(0)\rangle
      -\langle A_4(t-1)P(0)\rangle}{4\langle P(t)P(0)\rangle}\right) \, ,
\end{equation}
and in practice we can average over time separations satisfying
$12\leq t\leq20$.
Figure \ref{fig:pcacfit} shows the example of $\beta=2.2$ and $m_0=-0.75$.
The other additional quantity is the Goldstone boson decay constant that we
extract from a simultaneous fit to three
correlators \cite{deDivitiis:1995pf},
\begin{eqnarray}
\langle A_4(t)A_4(0)\rangle &=& \sum_{j=1}^n\frac{2m_j}{\sqrt{L^3}}
   \left(\frac{f_j}{Z_A}\right)^2\cosh\left(m_j\left(t-\frac{T}{2}\right)
   \right) \, , \label{fPieq1} \\
\langle A_4(t)P(0)\rangle &=& \sum_{j=1}^n\left(\frac{m_j}{2m_qZ_P}\right)
   \frac{2m_j}{\sqrt{L^3}}\left(\frac{f_j}{Z_A}\right)^2\cosh\left(m_j\left(
   t-\frac{T}{2}\right)\right) \, , \\
\langle P(t)P(0)\rangle &=& \sum_{j=1}^n\left(\frac{m_j}{2m_qZ_P}\right)^2
   \frac{2m_j}{\sqrt{L^3}}\left(\frac{f_j}{Z_A}\right)^2\cosh\left(m_j\left(
   t-\frac{T}{2}\right)\right) \, , \label{fPieq3}
\end{eqnarray}
where $L=16$ and $T=32$.  As stated above, we use
$n=3$ and fit all time separations except $t=0$.
The fitting parameters are $m_j$, $m_j/(m_qZ_P)$,
and $f_j/Z_A$.  Notice that such fits only provide the decay constant
divided by its renormalization constant $Z_A$.  Since $Z_A$ will approach
unity in the continuum limit, we will study the ratio $f_j/Z_A$ rather than
$f_j$ itself in this exploratory study.

\section{Results from Numerical Simulations}

\begin{figure}
\includegraphics[width=17cm,trim=0 50 0 0,clip=true]{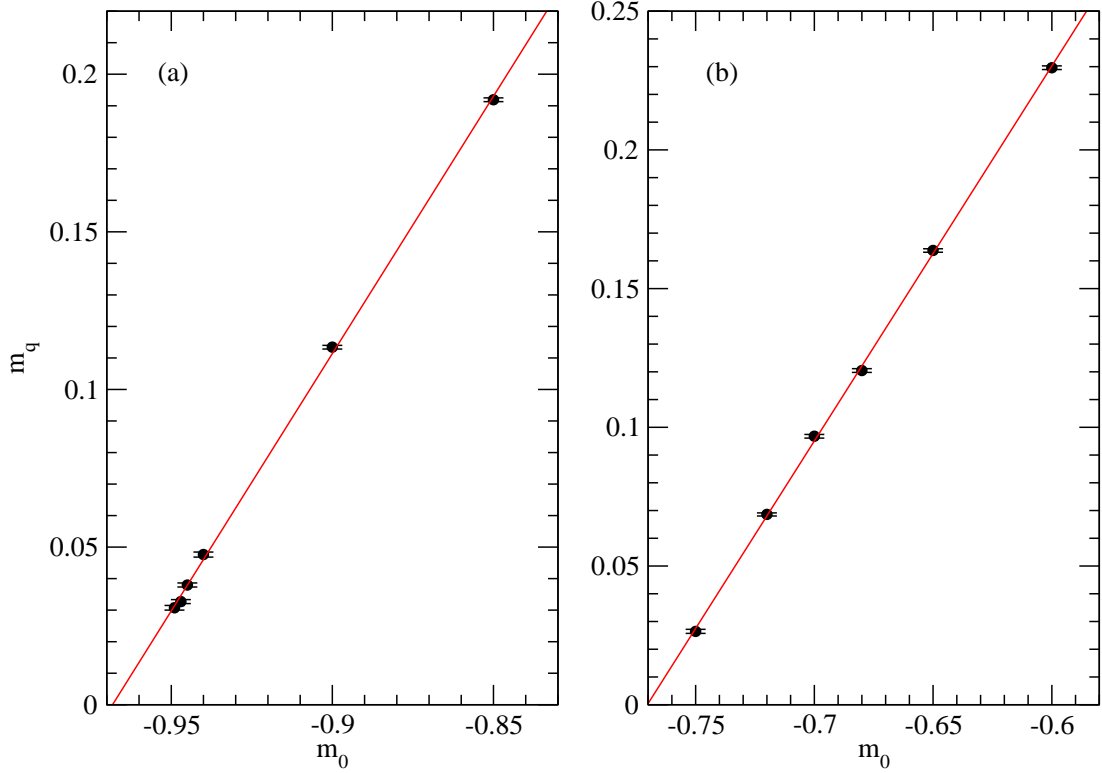}
\caption{The PCAC quark mass as a function of the bare quark mass for (a)
         $\beta=2.0$ and (b) $\beta=2.2$.  In each case, the line is a
         fit to all six data points.}\label{fig:pcacmass}
\end{figure}

As familiar from QCD, the axial Ward-Takahashi identity incorporates a
partially-conserved axial vector current (PCAC) which defines a renormalized
quark mass in terms of the axial current's derivative.  Use of the explicit
definition in (\ref{mqformula}) produces the relationship between
renormalized and bare quark masses as plotted in Fig.~\ref{fig:pcacmass}.
Over our range of bare masses, the PCAC mass is seen to be effectively linear
in the bare mass.

\begin{figure}
\includegraphics[width=17cm,trim=0 50 0 0,clip=true]{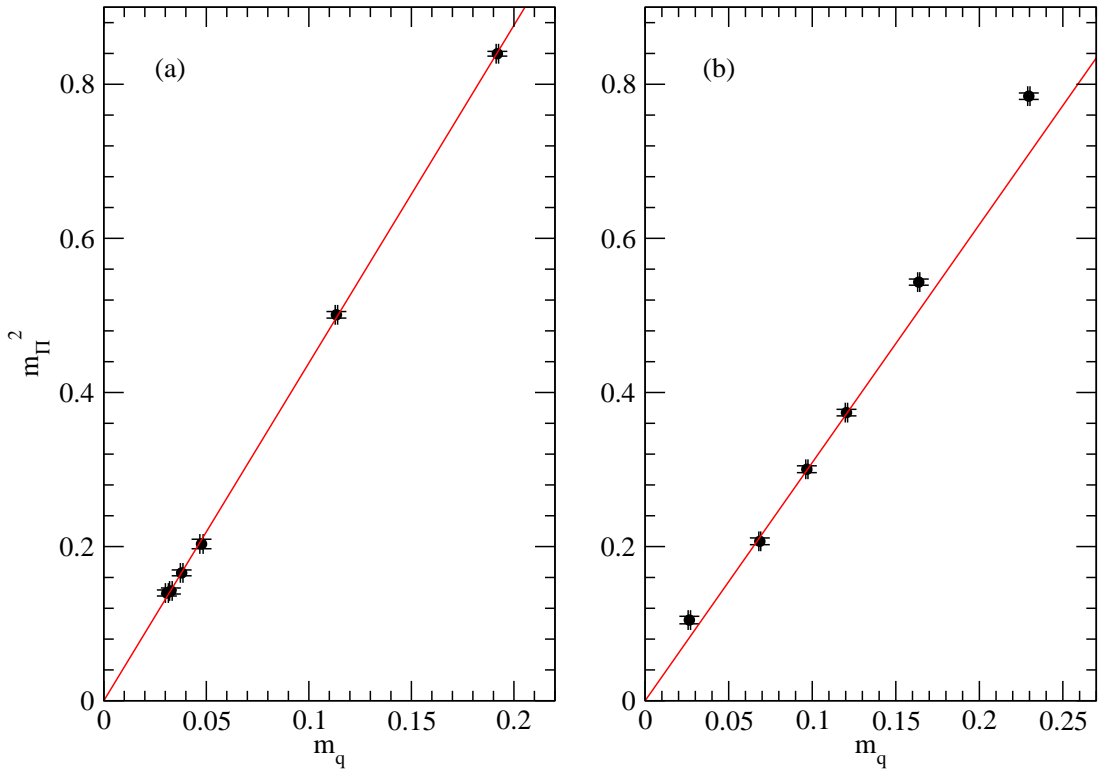}
\caption{The Goldstone boson mass squared as a function of the PCAC quark mass
         for (a)
         $\beta=2.0$ and (b) $\beta=2.2$.  In (a) the line is a one-parameter
         fit to all six data points, but in (b) the line is a fit to the
         three points within $0.05<m_q<0.15$.}\label{fig:GBmass}
\end{figure}

Goldstone bosons should be massless when the renormalized quark mass is
zero, and should approach that limit according to
\begin{equation}\label{leadingsqrt}
m_\Pi \propto \sqrt{m_q} \, .
\end{equation}
The lattice data confirm this behavior, as shown in Fig.~\ref{fig:GBmass}.
At $\beta=2.0$ our entire range of masses satisfies (\ref{leadingsqrt}).
At $\beta=2.2$ our range of mass values is broad enough that the two heaviest
quarks show curvature coming from corrections to (\ref{leadingsqrt}), and our
lightest quark displays a finite-volume correction.
We wish to emphasize that these plots in Fig.~\ref{fig:GBmass}
are obtained using any of the
five operators from (\ref{GBmesonops}) and (\ref{GBdiquarkops}).  All five
Goldstone bosons are exactly degenerate.  No single-site propagators were
needed for computing these correlations functions.

\begin{figure}
\includegraphics[width=17cm,trim=0 50 0 0,clip=true]{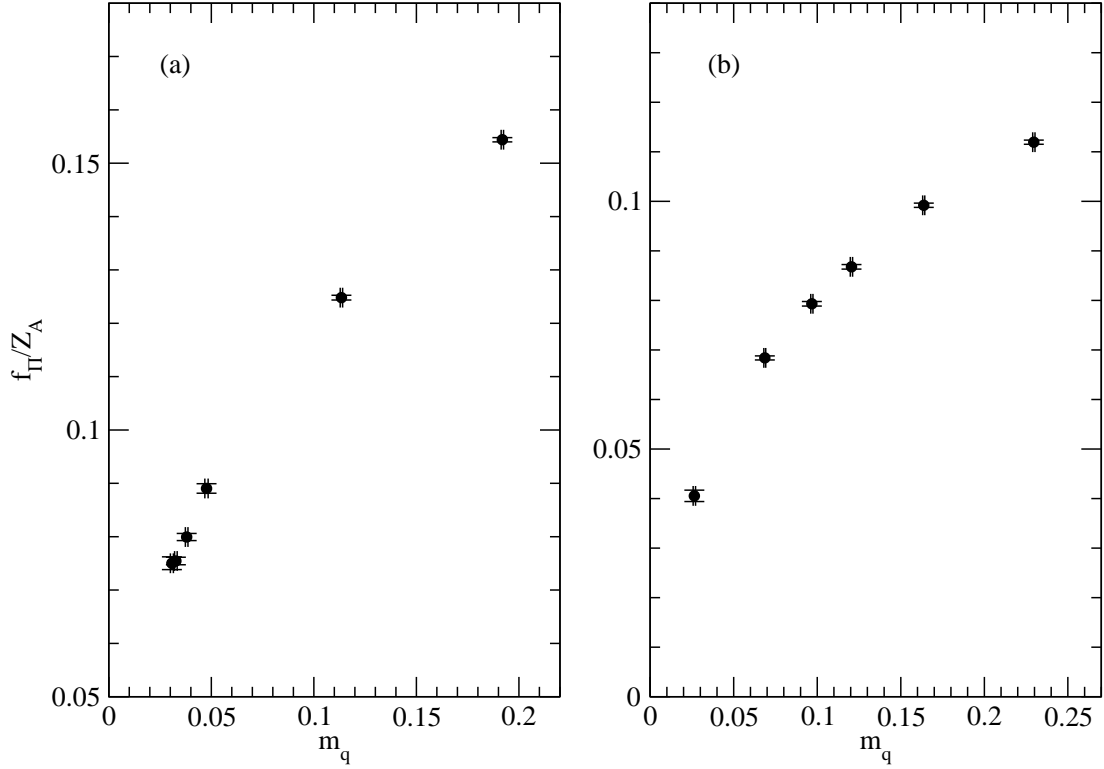}
\caption{The Goldstone boson decay constant as a function of the PCAC quark
         mass for (a) $\beta=2.0$ and (b) $\beta=2.2$.}\label{fig:fPi}
\end{figure}

The physical scale of technicolor can be specified by requiring that the
Goldstone decay constant matches the observed electroweak energy scale,
\begin{equation}
f_\Pi \approx 246 \, {\rm GeV.}
\end{equation}
Comparison to a lattice determination of $f_\Pi$ would thus provide a direct
interpretation of the physical scale for each lattice simulation.
Up to the renormalization constant $Z_A$, we can produce these lattice
results by using (\ref{fPieq1}-\ref{fPieq3}), as displayed in
Fig.~\ref{fig:fPi}.  A curvature is clearly visible in these plots of $f_\Pi/Z_A$
versus $m_q$, and for our purposes it is sufficient to notice that
$f_\Pi/Z_A\sim O(0.1)$ in our simulations, i.e.\ an order of magnitude below
the lattice cut-off.

\begin{figure}
\includegraphics[width=17cm,trim=0 50 0 0,clip=true]{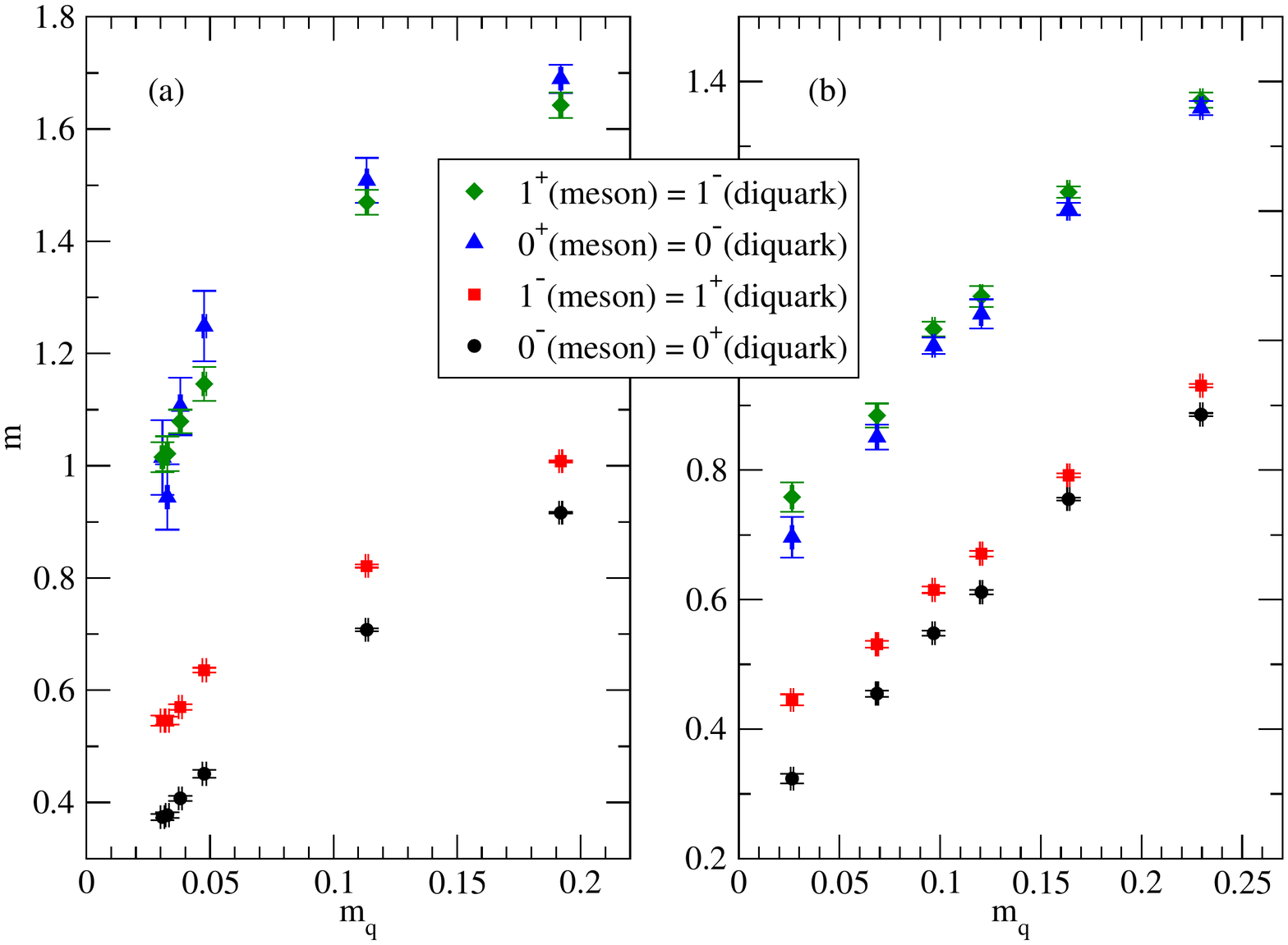}
\caption{Mass spectrum of isovector particles, with various $J^P$ quantum
         numbers, as a function of the PCAC quark
         mass for (a) $\beta=2.0$ and (b) $\beta=2.2$.}\label{fig:spectrum}
\end{figure}

{ Figure \ref{fig:spectrum}, besides displaying again the Goldstone isovector spectrum, can be used to verify that non-Goldstone type states, such as the scalar and spin one isovector particles, remain massive in the  $m_q=0$ limit.} Much like QCD, the vector meson mass is
nearly an order of magnitude larger than $f_\Pi$, but
significantly lighter than the isovector scalar and axial vector mesons.
Lattice artifacts are presumably sizable for masses $m\gtrsim1$, but our
results do extend into the region where all masses are less than $1$.
Notice the exact degeneracy between mesons and diquarks of opposite parity, as
anticipated in Section \ref{sec:EFT}.

Although isoscalar correlation functions bring the major challenge of
single-site propagators, we want to consider the specific case of the
isoscalar pseudoscalar meson.  Our goal is to verify explicitly that this meson
is not an additional Goldstone particle.
\begin{figure}
\includegraphics[width=17cm,trim=0 50 0 0,clip=true]{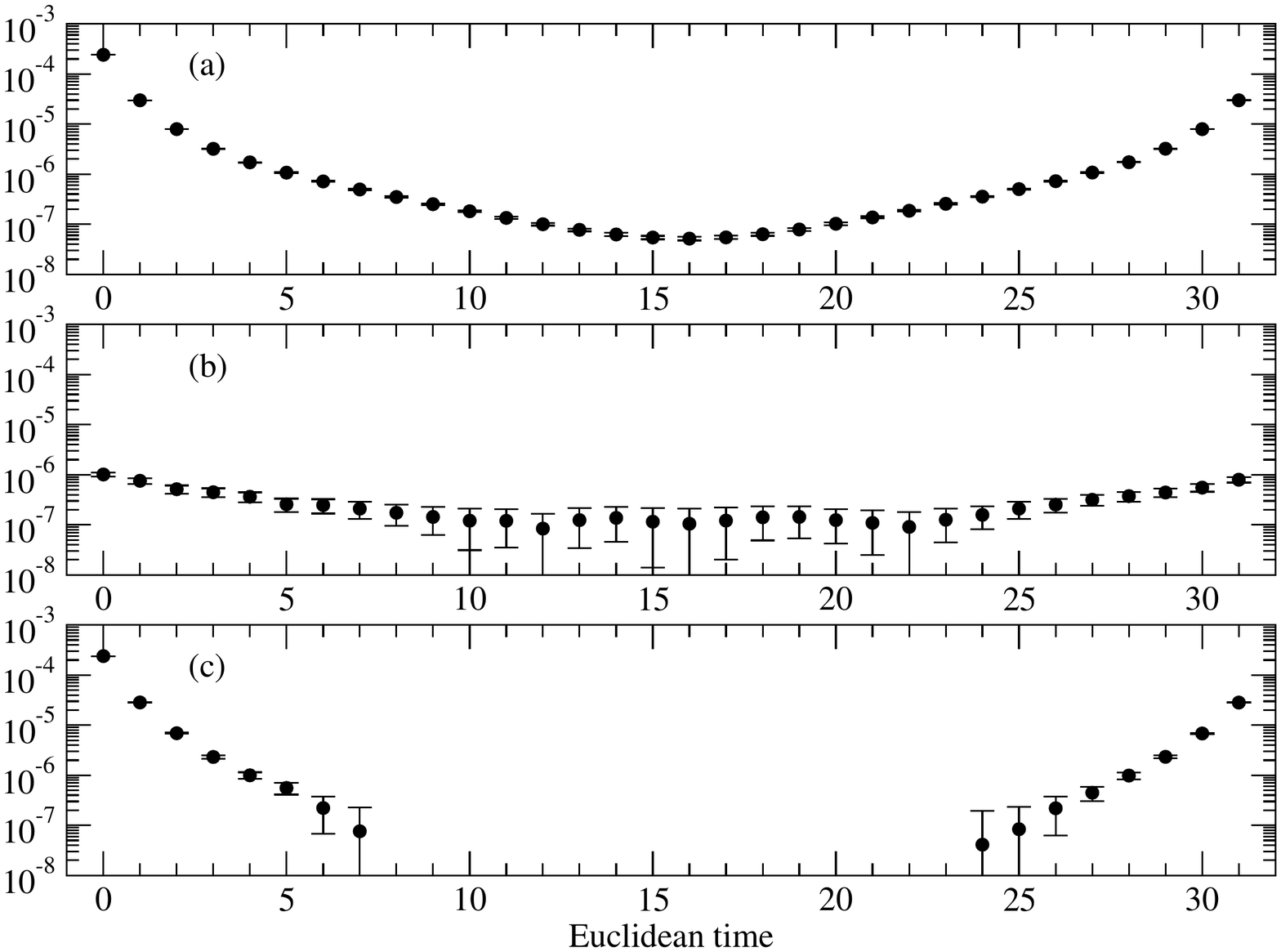}
\caption{The (a) connected, (b) disconnected, and (c) total contributions to
         the correlation function of an isoscalar pseudoscalar meson at
         $\beta=2.2$ and $m_0=-0.75$.}\label{fig:discon}
\end{figure}
Figure~\ref{fig:discon} displays the results.  Panel (a) is the correlation
function for the isovector, i.e.\ the quark-connected diagram only.
A multi-exponential fit to those data produce the corresponding data point
in Fig.~\ref{fig:GBmass}.
Panel (b) is the contribution from the quark-disconnected diagram built from
a pair of single-site quark propagators.
Panel (c) is obtained from the upper two panels according to ``(c)=(a)-2(b)'',
where the factor of 2 accounts for $\overline{U}U$ plus $\overline{D}D$
contributions and the minus sign accounts for anticommuting fermions.
In panel (c), data at all timesteps between 7 and 24 are statistically
consistent with zero, and time steps near the source at
$t=0$ correspond to states much heavier than the Goldstones.  Thus we see how
the quark-disconnected diagrams have canceled the Goldstone signal
out of the isoscalar.  This is the expected conclusion since the symmetry
breaking $SU(4)\to Sp(4)$ requires exactly five Goldstone bosons.

\section{Conclusions} 

From many previous studies \cite{Nussinov:1985xr,Gudnason:2006ug,Foadi:2008qv,Khlopov:2008ty,Dietrich:2006cm,Sannino:2009za,Ryttov:2008xe,Kaplan:2009ag,Frandsen:2009mi,Belyaev:2010kp}
the lightest neutral technibaryon has become a classic prototype for
asymmetric DM,
and \cite{Gudnason:2006ug,Foadi:2008qv,Ryttov:2008xe,DelNobile:2011je} focused on the possibility of that technibaryon
being a pseudo-Goldstone boson.  In this work, we have performed lattice
simulations of $SU(2)$ gauge theory with two techniquarks which has been used as a template for the construction of composite light DM models  \cite{Ryttov:2008xe,DelNobile:2011je}. For the first time, we verify the existence of the pseudo-Goldstone state to identify with the composite DM candidate. In fact, besides the standard three Goldstone bosons, to be eaten up by the SM gauge bosons, we  observe two more Goldstone technibaryons. These latter states are scalars rather than pseudoscalars.
The masses of other phenomenologically relevant technihadrons were found to be an order of magnitude
heavier than the Goldstone decay constant, thus justifying the construction of
an effective field theory containing only the five Goldstone fields introduced in \cite{Appelquist:1999dq}  and then adapted to UMT in \cite{Ryttov:2008xe}. We have established the expected pattern of chiral symmetry breaking in the absence of the electroweak interactions. Typically these interactions tend to destabilize the vacuum (see \cite{Ryttov:2011my} for a recent study) which can be, however, re-stabilized when extending the technicolor theory to give masses to the SM fermions and to the otherwise massless DM candidate. 

One might ask whether a light isoscalar scalar meson is present in this gauge
theory, reminiscent of the $\sigma/f_0(600)$ meson of QCD, that could be viewed
as a composite Higgs boson.  The isoscalar scalar channel is a major
challenge for lattice simulations \cite{Kunihiro:2009zz}, and we must leave this
question to be the focus of a future study.

In the future it will be very interesting to determine more accurately
the mass spectrum of the theory, but there are many additional topics
waiting to be explored as well.  Here we mention three of them.
The effects of
{\em four-fermion interactions}, relevant for a realistic extension of the
SM, should be investigated \cite{Fukano:2010yv}.  The
extension to {\em nonzero temperature} would lead to a characterization of
the early universe electroweak phase transition.  Knowledge of its order and
strength is essential for investigating the possibility of providing correct
baryon and asymmetric DM genesis at the electroweak scale
\cite{Cline:2008hr,Jarvinen:2009wr,Jarvinen:2009pk}.  The {\em nonzero matter
density} regime is also useful for asymmetric DM physics
\cite{Gudnason:2006ug,Ryttov:2008xe} given that an asymmetry implies the
presence of a nonzero chemical potential.  In contrast to QCD, the extension
to nonzero matter density is directly possible in lattice simulations of this
$SU(2)$ theory.  That is because the fermions belong to a pseudoreal
representation of the $SU(2)$ gauge group and therefore a chemical potential
does not lead to an imaginary action.  At nonzero matter density a number of
interesting phases can emerge, such as baryon superfluidity
(due to the condensation of the diquark technibaryon) and
rotational invariance breaking (due to spin-one condensates
\cite{Sannino:2002wp,Lenaghan:2001sd,Sannino:2001fd} arising from Lorentz
symmetry breaking at nonzero chemical potential).

\acknowledgments

This work was supported in part by the Natural Sciences and Engineering
Research Council (NSERC) of Canada.  Computing facilities were provided by
the Shared Hierarchical Academic Research Computing Network
(SHARCNET: http://www.sharcnet.ca).

\appendix
\section{Proof of Equation \eqref{14}}
\label{Appendix1}
The Wilson fermion matrix as obtained from Eq.~\eqref{wilsonm} is
\begin{equation}
\langle U(y)\overline{U}(x) \rangle^{-1} = (4+m_0)\delta_{xy}
    - \frac{1}{2}\sum_\mu\bigg((1-\gamma_\mu) U_\mu(y)\delta_{y+\mu,x}
      +(1+\gamma_\mu) U_\mu^\dagger(x)\delta_{y,x+\mu}\bigg) \, .
\label{1A}
\end{equation}
The transpose is
\begin{equation}
\langle [(U(y)\overline{U}(x)]^T \rangle^{-1} = (4+m_0)\delta_{xy}
    - \frac{1}{2}\sum_\mu\bigg((1-\gamma_\mu^T) U_\mu^T(y)\delta_{y+\mu,x}
      +(1+\gamma_\mu^T) U_\mu^*(x)\delta_{y,x+\mu}\bigg) \, .
\end{equation}
Applying the charge conjugation operator $C$ we obtain
\begin{equation}
C^{-1} \langle [U(y)\overline{U}(x)]^T \rangle^{-1} C = (4+m_0)\delta_{xy}
    - \frac{1}{2}\sum_\mu\bigg((1+\gamma_\mu) U_\mu^T(y)\delta_{y+\mu,x}
      +(1-\gamma_\mu) U_\mu^*(x)\delta_{y,x+\mu}\bigg) \, .
\end{equation}
 Recalling that for a general $SU(2)$ matrix ${\cal U}$ the following identity holds:
 \begin{eqnarray*}
(-i\sigma^2)^{-1} {\cal U} (-i\sigma^2)
 = {\cal U}^* \, ,
\end{eqnarray*}
we deduce:
\begin{eqnarray}
&& C^{-1}(-i\sigma^2)^{-1} \langle [U(y)\overline{U}(x)]^T  \rangle^{-1} C(-i\sigma^2)
    \nonumber \\
&=& (4+m_0)\delta_{xy} - \frac{1}{2}\sum_\mu\bigg(
    (1+\gamma_\mu) U_\mu^\dagger(y)\delta_{y+\mu,x}
   +(1-\gamma_\mu) U_\mu(x)\delta_{y,x+\mu}\bigg) \, .
\end{eqnarray}
By direct comparison with Eq.~\eqref{1A}  we arrive at
\begin{equation}
C^{-1}(-i\sigma^2)^{-1} \langle [U(y)\overline{U}(x)]^T \rangle C(-i\sigma^2)
= \langle U(x)\overline{U}(y) \rangle \ ,
\end{equation}
which corresponds to Eq.~\eqref{14}. In the main text we dropped the $\langle .. \rangle$ symbols, indicating the vacuum expectation value, to ease the notation.

\end{document}